\documentclass[aps,prb,twocolumn,superscriptaddress,showpacs,amsmath,amssymb,footinbib,longbibliography,10pt]{revtex4-2}
\usepackage[utf8]{inputenc}
\usepackage{bm}
\usepackage[dvipsnames]{xcolor}
\usepackage{multirow}
\usepackage{amssymb}
\usepackage{amsbsy}
\usepackage{amsmath}
\usepackage{stmaryrd}
\usepackage{graphicx}
\usepackage{epsfig}
\usepackage{placeins}
\usepackage[normalem]{ulem}
\usepackage{bbold}
\usepackage{bbm}
\usepackage{braket}
\usepackage{graphicx}
\usepackage{tikz}
\usetikzlibrary{calc}
\usetikzlibrary{patterns}
\usepackage[colorlinks,linkcolor=blue,citecolor=blue,urlcolor=blue]{hyperref}

\makeatletter

\setcitestyle{numbers,square,comma,sort&compress}

\definecolor{palette1}{HTML}{A8216B}
\definecolor{palette2}{HTML}{F1184C}
\definecolor{palette3}{HTML}{F36943}
\definecolor{palette4}{HTML}{F7DC66}
\definecolor{palette5}{HTML}{2E9599}

\usepackage{scalerel}
\usepackage{tikz}
\usetikzlibrary{calc}
\usetikzlibrary{patterns}
\usetikzlibrary{svg.path}
\definecolor{orcidlogocol}{HTML}{A6CE39}
\tikzset{
  orcidlogo/.pic={
    \fill[orcidlogocol]
svg{M256,128c0,70.7-57.3,128-128,128C57.3,256,0,198.7,0,128C0,57.3,57.3,0,128,
0C198.7,0,256,57.3,256,128z};
    \fill[white] svg{M86.3,186.2H70.9V79.1h15.4v48.4V186.2z}

svg{M108.9,79.1h41.6c39.6,0,57,28.3,57,53.6c0,27.5-21.5,53.6-56.8,53.6h-41.8V79.
1z
M124.3,172.4h24.5c34.9,0,42.9-26.5,42.9-39.7c0-21.5-13.7-39.7-43.7-39.7h-23.
7V172.4z}

svg{M88.7,56.8c0,5.5-4.5,10.1-10.1,10.1c-5.6,0-10.1-4.6-10.1-10.1c0-5.6,4.5-10.1
,10.1-10.1C84.2,46.7,88.7,51.3,88.7,56.8z};
  }
}

\newcommand\orcid[1]{\!%
  \href{https://orcid.org/#1}{%
    \mbox{%
      \scaleto{%
        \begin{tikzpicture}[yscale=-1,transform shape]
          \pic{orcidlogo};
        \end{tikzpicture}
      }{8pt}%
    }%
  }%
}

\begin{document}
\title{Lindblad quantum dynamics from correlation functions of classical spin
chains}

\author{Markus Kraft~\orcid{0009-0008-4711-5549}}
\email{markus.kraft@uos.de}
\affiliation{University of Osnabr{\"u}ck, Department of Mathematics/Computer
Science/Physics, D-49076 Osnabr{\"u}ck, Germany}

\author{Mariel Kempa~\orcid{0009-0006-0862-4223}}
\affiliation{University of Osnabr{\"u}ck, Department of Mathematics/Computer
Science/Physics, D-49076 Osnabr{\"u}ck, Germany}

\author{Jiaozi Wang~\orcid{0000-0001-6308-1950}}
\affiliation{University of Osnabr{\"u}ck, Department of Mathematics/Computer
Science/Physics, D-49076 Osnabr{\"u}ck, Germany}

\author{Robin Steinigeweg~\orcid{0000-0003-0608-0884}}
\email{rsteinig@uos.de}
\affiliation{University of Osnabr{\"u}ck, Department of Mathematics/Computer
Science/Physics, D-49076 Osnabr{\"u}ck, Germany}

\date{\today}


\begin{abstract}
The Lindblad quantum master equation is one of the central approaches to the
physics of open quantum systems. In particular, boundary driving enables the
study of transport, where a steady state emerges in the long-time limit, which
features a constant current and a characteristic density profile. While the
Lindblad equation complements other approaches to transport in closed quantum
systems, it has become clear that a connection between closed and open systems
exists in certain cases. Here, we build on this connection for magnetization
transport in the spin-1/2 XXZ chain with and without integrability-breaking
perturbations. Specifically, we study the question whether the
time evolution of the open quantum system can be described on the basis of
classical correlation functions, as generated by the Hamiltonian equations of
motion for real vectors. By comparing to exact numerical simulations of the
Lindblad equation, we find a good accuracy of such a
description for a range of model parameters, which is
consistent with previous studies on closed systems. While this
agreement is an interesting physical observation, it also suggests that
classical mechanics can be used to solve the Lindblad equation for
comparatively large system sizes, which lie outside the possibilities of a
quantum mechanical treatment. We also point out counterexamples
and limitations for the quantitative extraction of transport coefficients.
Remarkably, our classical approach to large open systems allows
to detect superdiffusion at the isotropic point.
\end{abstract}

\maketitle


\section{Introduction}

Many-body quantum systems out of equilibrium can be studied in two complementary
scenarios. Either, the system is closed and couples by no means to the rest
of the world. Or, the system is open and explicitly couples to a bath, where
the system-bath coupling can be weak or strong. Both scenarios are interesting
and important by their own right and allow to address a large variety of
questions in modern physics, ranging from fundamental questions in statistical
mechanics to applied questions in material science. A central question in
closed and open scenarios is the system's evolution in the course of time and
the existence and properties of steady states in the long-time limit
\cite{Polkovnikov2011, Eisert2015, Dalessio2016, Borgonovi2016, Abanin2019}.
The study of this particular question has seen remarkable progress in the past,
due to experimental advances \cite{Bloch2008}, fresh theoretical concepts
like typicality of pure quantum states \cite{Gemmer2004, Goldstein2006,
Popescu2006, Reimann2007, Bartsch2009, Elsayed2013, Steinigeweg2014,
Heitmann2020, Jin2021} and eigenstate thermalization \cite{Deutsch1991,
Srednicki1994, Rigol2008}, and the development of sophisticated numerical
techniques \cite{Schollwoeck2005, Schollwoeck2011}.

Within the diverse class of nonequilibrium processes, transport is a natural
one for systems with one or more globally conserved quantities
\cite{Bertini2021}, like total energy, particle number, or magnetization.
Transport is further a process which is relevant to both, closed and open
situations. In an open situation, the system of actual interest can be coupled
at its boundaries to two reservoirs at different temperatures or chemical
potentials, such that transport is induced and a nonequilibrium steady state is
usually established in the long-time limit \cite{Michel2003, Prosen2009,
Znidaric2011}. In this steady state, the constant current and the form
of the density profile yield information on the qualitative type of transport
and also allow to determine quantitative values for transport coefficients
\cite{,Bertini2021}. A widely used strategy to describe such an open scenario
is the Lindblad equation \cite{Breuer2007}, which has its assets and drawbacks.
On the one hand, the derivation of this equation from a microscopic system-bath
model can be a nontrivial task in praxis \cite{Wichterich2007, DeRaedt2017}. On
the other hand, it is the most general form of a quantum master equation, which
Markovian and maps any density matrix to a density
matrix again. Furthermore, the structure of the Lindblad equation enables the
application of well-suited numerical techniques. Here, one method is provided
by the concept of stochastic unraveling \cite{Dalibard1992, Michel2008}, which
constructs the time evolution of the density matrix in Liouville space as the
average over many pure-state trajectories in Hilbert space. An alternative
method is given by matrix product states \cite{Zwolak2004, Verstraete2008,
Prosen2009, Weimer2021}, where entanglement growth is reduced because of
dissipation.

In closed systems, a main approach is linear response theory, which predicts
the behavior close to equilibrium in terms of correlation functions at
equilibrium. While in the context of transport the current autocorrelation
is a central object and enters the well-known Kubo formula \cite{Kubo1991}, the
transport behavior is also encoded in density-density correlations, which can be
analyzed in real or momentum space and in the time or frequency domain. Even
though the investigation of correlation functions has a long and fertile
history, the concrete calculation for specific models can still be a
challenging task in praxis. In particular, seemingly simple models like the
integrable spin-$1/2$ XXZ chain have turned out to be notoriously difficult for
both, analytical and numerical methods \cite{Bertini2021}. Models of
interacting spins are valuable, because they are not only many-body quantum
systems with rich phase diagrams, but also have a classical counterpart
\cite{Windsor1967, Huber1969, Lurie1974, DeRaedt1981, Mueller1988, Gerling1989,
Gerling1990, AlcantaraBonfim1992, Boehm1993, Srivastava1994, Constantoudis1997,
Oganesyan2009, Huber2012, Steinigeweg2012, Wijn2012, Bagchi2013, Prosen2013a,
Jin2013, Jencic2015, Das2018, Das2019, Das2019a, Li2019, Glorioso2021,
McRoberts2022, Roy2023a, Benet2023, McRoberts2024}, which corresponds to the
limit of large spin quantum numbers $S \to \infty$. While the cases of $S=1/2$
and $S \to \infty$ can in general not be expected to exhibit the same physics,
it has been observed that their dynamics is similar for some examples
\cite{Elsayed2015, Gamayun2019}, qualitatively and also quantitatively. Such a
similarity has been found for correlation functions in the XXZ chain in the
limit of high temperatures $T \to \infty$ \cite{Schubert2021, Heitmann2022a},
where the low-energy excitations are less relevant \cite{Park2024}.

In this paper, we also investigate to what extent the time evolution in a
quantum system can be described on the basis of the dynamics in the classical
counterpart. In contrast to previous works, which have been devoted to a
comparison of the corresponding correlation functions in a closed scenario
\cite{Schubert2021, Heitmann2022a}, we intend to go a substantial step beyond.
Specifically, we are going to compare the open quantum system to the closed
classical system. For such a comparison, we obviously need a connection between
the dynamics in open and closed scenarios \cite{Steinigeweg2009a,
Znidaric2019}, which does not exist in general \cite{Kundu2009,
Purkayastha2018, Purkayastha2019}. For the spin-$1/2$ chain, however, a
connection has been recently suggested \cite{Heitmann2023} for the case of
small system-bath coupling and weak driving. Because this connection involves
quantum correlation functions, we replace them by the corresponding classical
ones. In this way, we can address the main question of our work: Is it possible
to obtain Lindblad quantum dynamics from correlation functions of classical
spin chains? To answer this question, we compare to exact numerical simulations
of the Lindblad equation. We observe a good agreement for a range of model
parameters, which is consistent with previous studies on closed
systems. This agreement further hints that classical mechanics can be used as
a strategy to solve the Lindblad equation for large system sizes, which are not
accessible in a quantum mechanical treatment. Such a classical strategy has
been also discussed in other open quantum systems \cite{Hogg2024}.
We also point out counterexamples and limitations for the
quantitative extraction of transport coefficients. Remarkably,
our classical approach to large open systems allows to detect
superdiffusion at the isotropic point.

This paper is structured as follows: We introduce the open quantum system in
Sec.\ \ref{sec:open}. Then, we discuss the connection between open and closed
systems in Sec.\ \ref{sec:prediction} and the classical limit in Sec.\
\ref{sec:classical}. Next, we present our numerical results in Sec.\
\ref{sec:results}. We close with a summary and conclusion in Sec.\
\ref{sec:conclusion}. Further information can be found in the appendix.


\section{Open quantum system} \label{sec:open}

\begin{figure}[t]
\includegraphics[width=0.9\columnwidth]{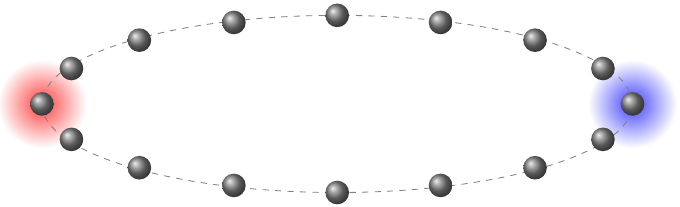}
\caption{Geometry of the open quantum system. The
spin-$1/2$ XXZ chain with
periodic boundary conditions is coupled to two Lindblad baths, which are
located at a site $B_{1}=1$ and another site $B_{2}= N/2 +1$.}
\label{fig:sketch}
\end{figure}

Let us start by introducing the open quantum system considered throughout this
work. We choose to describe this system by the Lindblad equation,
\begin{equation} \label{eq:Lindblad}
\dot{\rho}(t) = \mathcal{L}\rho(t) = i [\rho(t),H] + \mathcal{D} \rho(t)
\; ,
\end{equation}
as the most general form of a quantum master equation, which is
Markovian
and maps any density matrix to a density matrix again \cite{Breuer2007}. While
the first term on the r.h.s.\ is coherent and describes the unitary time
evolution w.r.t.\ to a given Hamiltonian $H$, the second term on the r.h.s.\ is
incoherent and describes the damping due to the presence of a bath. This
damping reads
\begin{equation} \label{eq:damping}
\mathcal{D} \rho(t) = \sum_{j} \alpha_{j} \left( L_{j}\rho(t)L_{j}^{\dagger} -
\frac{1}{2} \left \{  \rho(t),L_{j}^{\dagger}L_{j} \right \} \right) \,
\end{equation}
with Lindblad operators $L_{j}$, non-negative rates $\alpha_{j}$, and the
anticommutator $\{\bullet, \bullet \}$.

Next, we define the Hamiltonian $H$ and a suitable set of Lindblad
operators $L_{j}$. Focusing first on $H$, we choose the spin-$1/2$ XXZ
chain \cite{Bertini2021},
\begin{equation} \label{eq:Hamiltonian_spin}
H = J \sum_{r=1}^{N} \left ( S_{r}^{x} S_{r+1}^{x} + S_{r}^{y} S_{r+1}^{y} +
\Delta S_{r}^{z}S_{r+1}^{z} \right ) \; ,
\end{equation}
where $S_{r}^{j}$ $(j = x,y,z)$ are the components of a spin-$1/2$ operator at
site $r$, $N$ is the number of sites, $J > 0$ is the antiferromagnetic
coupling constant, and $\Delta$ is the anisotropy in $z$ direction.
We assume periodic boundary conditions, $S_{N+1}^{j}
=S_{1}^{j}$, which are for our purposes more convenient than open boundary
conditions, as discussed later.

Because the Hamiltonian in Eq.\
(\ref{eq:Hamiltonian_spin}) is integrable for all possible values of the
anisotropy $\Delta$, we additionally take into account an
integrability-breaking perturbation. Our choice for such a perturbation are
interactions between next-to-nearest sites,
\begin{equation} \label{eq:Hamiltonian_spin_perturbation}
H' = H + \Delta^{\prime} \sum_{r=1}^{N} S_{r}^{z}S_{r+2}^{z} \; ,
\end{equation}
where $\Delta^{\prime}$ is the perturbation strength and, as before, we
assume periodic boundary conditions.

For each of the Hamiltonians in Eqs.\ (\ref{eq:Hamiltonian_spin}) and
(\ref{eq:Hamiltonian_spin_perturbation}), the total magnetization $S^{z} =
\sum_{r} S_{r}^{z}$ is a strictly conserved quantity, $[H,S^{z}] = [H',S^{z}] =
0$. Thus, the transport of local magnetization is a meaningful
question. The study of transport also motivates our choice of the Lindblad
operators $L_{j}$. Specifically, we make a simple but common choice
\cite{Bertini2021},
\begin{align}
L_{1} = S^{+}_{B_{1}}, &\ \; \alpha_{1} = \gamma(1+\mu) \, , \label{eq:L1} \\
L_{2} = L^{\dagger}_{1} = S^{-}_{B_{1}}, &\ \; \alpha_{2} = \gamma(1-\mu) \,
, \\
L_{3} = S^{+}_{B_{2}}, &\ \; \alpha_{3} = \gamma(1-\mu) \, ,\\
L_{4} = L^{\dagger}_{3} = S^{-}_{B_{2}}, &\ \; \alpha_{4} = \gamma(1+\mu) \,
, \label{eq:L4}
\end{align}
where $\gamma$ is the system-bath coupling and $\mu$ is the driving strength.
The local operators $L_{1}$ and $L_{2}$ act on the site $B_{1}$ and flip a spin
up and down, respectively. The other operators $L_{3}$ and $L_{4}$ act in the
same way on another site $B_{2}$. To maximize the size of the bulk,
we set $B_{1}=1$ and $B_{2}= N/2 +1$, as illustrated in Fig.\ \ref{fig:sketch}.
Due to the choice of the rates in Eqs.\ (\ref{eq:L1}) - (\ref{eq:L4}),
net magnetization flows from the first bath into the system and from the system
into the second bath, which leads to a nonequilibrium steady state in the
long-time limit.

For this open quantum system, we are interested in the dynamics of local
magnetization, which includes the steady-state profile on the one hand and
its buildup in time on the other hand. Hence, we study the expectation value
\begin{eqnarray} \label{eq:exp_value_magnetization}
\langle S_{r}^{z}(t) \rangle = \text{tr}[S_{r}^{z}\rho(t)] \; ,
\end{eqnarray}
which depends on the Hamiltonian $H$, but also on the system-bath coupling
$\gamma$ and the driving strength $\mu$. We focus on the case of small $\gamma$
and $\mu$. As initial condition, we use the ensemble
\begin{equation}
\rho(0) = \frac{e^{-\beta H}}{\text{tr}[e^{-\beta H}]}
\end{equation}
for high temperatures $\beta = 1/T \to 0$, which features a homogeneous profile
of magnetization.

An exact analytical solution of the Lindblad equation, or an accurate
approximation of it, can in general not be derived. Hence, one has to
resort typically to numerical methods. In this context, standard
exact
diagonalization is particularly challenging \cite{SU1}, since the Liouville
space (of
dimension $D = 2^N \times 2^N)$ is substantially larger than the anyhow large
Hilbert space ($D = 2^N$). Yet, the Lindblad form allows for stochastic
unraveling \cite{Dalibard1992, Michel2008}, which yields the time-dependent
density matrix as the average over many pure-state trajectories. Additionally,
simulations on the basis of matrix product states \cite{Zwolak2004,
Verstraete2008, Prosen2009, Weimer2021} give access to systems of hundreds of
spins, at least on time scales with a still low amount of entanglement.

In this work, we mostly rely on existing numerical data in the literature
\cite{Heitmann2023, Kraft2024},
which we use later for a comparison to our approach on the basis of classical
mechanics. Before we explain what we mean by classical mechanics, we need to
discuss another concept, i.e., a connection between the dynamics in open and
closed quantum systems.


\section{Connection between open and closed systems} \label{sec:prediction}

In general, one can hardly expect a direct connection between the time
evolution in open and closed quantum systems. For the specific scenario
introduced in Sec.\ \ref{sec:open}, however, such a connection has been shown
to exist, at least in certain cases \cite{Heitmann2023}. This
connection makes use of spatio-temporal correlation functions,
\begin{equation} \label{eq:corr_function_magnetization_quantum}
\langle S_{r}^{z}(t) S_{r'}^{z}(0)\rangle_{\text{eq}} =
\frac{\text{tr}\left[e^{-\beta H} e^{i H t} S_{r}^{z} e^{-i H t}
S_{r'}^{z}\right]}{\text{tr}\left [ e^{-\beta H}\right ]} \, ,
\end{equation}
which are evaluated in the closed system $H$ at thermal equilibrium. For
high temperatures $\beta = 1/T \to 0$, which we use from now on, they
simplify to
\begin{equation} \label{eq:corr_function_magnetization_quantum_simple}
\langle S_{r}^{z}(t) S_{r'}^{z}(0) \rangle_{\text{eq}} =
\frac{\text{tr}\left[e^{i H t} S_{r}^{z} e^{-i H t} S_{r'}^{z}\right]}
{2^{N}} \, .
\end{equation}
Before we formulate the actual connection, it is useful to introduce a
superposition of Eq.\ (\ref{eq:corr_function_magnetization_quantum_simple})
with $r' = B_1$ and Eq.\ (\ref{eq:corr_function_magnetization_quantum_simple})
with $r' = B_2$,
\begin{equation}
C_{r}(t) = \langle S_r^z(t) S_{B_1}^z(0) \rangle_\text{eq}  - \langle S_r^z(t)
S_{B_2}^z(0) \rangle_\text{eq} \, ,
\end{equation}
and then to define the more complex superposition
\begin{equation} \label{eq:trajectory}
d_{r}(t) = 2 \mu \sum_j A_j  \, \Theta(t -\tau_j) \, C_{r}(t-\tau_{j}) \, ,
\end{equation}
where $A_j$ are some amplitudes, $\tau_j$ are some times, and $\Theta(t)$ is
the Heavyside function. Using this notation, we can eventually formulate the
connection and express the nonequilibrium dynamics of the open system in Eq.\
(\ref{eq:exp_value_magnetization}) as \cite{Heitmann2023}
\begin{equation} \label{eq:prediction}
\langle S_{r}^{z}(t)\rangle \approx  \frac{1}{T_{\text{max}}}
\sum_{T=1}^{T_{\text{max}}} d_{r, T}(t) \, ,
\end{equation}
where the sum runs over $T_\text{max}$ different time sequences
$(\tau_1, \tau_2, \ldots)$. This sum over time
sequences gives rise to the additional index $T$, which is absent in Eq.\
(\ref{eq:trajectory}). Here, a particular time sequence is generated by
\begin{equation}
 \tau_{j+1} = \tau_{j} - \ln \frac{\varepsilon_{j+1}}{2\gamma} \, ,
\end{equation}
where $\varepsilon_{j+1}$ are random numbers drawn from a uniform distribution
in the  interval $]0,1]$. The amplitudes $A_j$, as derived in
\cite{Heitmann2023}, are now given by
\begin{equation} \label{eq:amplitues}
A_{j} = \frac{a_{j}\ -\ d_{B_{1},T}(\tau_{j} - 0^{+})}{\mu}
\end{equation}
with
\begin{equation}
a_{j} = \frac{\mu\ -\ 2\ d_{B_{1},T}(\tau_{j} - 0^{+})}{2\ -\ 4\mu\
d_{B_{1},T}(\tau_{j} - 0^{+})} \, .
\end{equation}
These amplitudes follow from $d_{B_{1},T}(\tau_{j} -
0^{+})$, which is evaluated at the bath-contact site $B_{1}$ and just
before the time $\tau_j$. Therefore, $A_j$ only depends on $A_{j-1}, \ldots,
A_1$. If $d_{B_1}(\tau_j - 0^{+}) \to 0$, $A_j \to 1/2$. Note that $A_1 = 1/2$
by construction. Loosely speaking, $A_j$ can be understood as the current
``boundary condition'' of the density profile, which should not be confused
with the actual boundary condition of the Hamiltonian.

While the connection might look quite complicated at first sight, it is
rather simple, especially since it expresses the nonequilibrium dynamics in
the open system just as a superposition of equilibrium correlation functions in
the closed system. As discussed in Ref.\
\cite{Heitmann2023} in detail, such a connection cannot always hold and
requires sufficiently small values of both, $\gamma$ and $\mu$.
Thus, one has to check in practise that the result for a given model and a
specific choice of the parameters $\gamma$ and $\mu$ does not change when these
parameters get smaller and smaller.

A sophisticated criterion for the smallness of the two
parameters also exists and has been studied in Ref.\ \cite{Kraft2024} in
detail. Loosely speaking, the life time of an excitation, which is injected at
the boundary of the model, has to be short compared to the time scale
$1/\gamma$. More formally, the autocorrelation function at the bath-contact
site,
\begin{equation}
\langle S_{B_1}^{z}(t) S_{B_1}^{z}(0) \rangle_{\text{eq}}
\end{equation}
has to decay on this time scale. For the specific case of periodic boundary
conditions,
which are used throughout our work, this condition can be satisfied by a
sufficiently small $\gamma$ and hence long enough $1/\gamma$. For the other
case of open boundary conditions, however, the autocorrelation function does
not decay at all for $\Delta > 1$, see Ref.\ \cite{Kraft2024} and also Refs.\
\cite{Fendley2016,
Kemp2017}.

To understand the sophisticated criterion for $\mu$, one has to resort to
stochastic unraveling, which underlies the derivation of the connection in Eq.\
(\ref{eq:prediction}) \cite{Heitmann2023}. In this context, $\mu$ has then to be
negligible for the deterministic evolution up to the average jump time
$1/\gamma$. For the jump probabilities, however, $\mu$ is fully taken into
account and gives rise to the amplitues in Eq.\ (\ref{eq:amplitues}).
Alternatively, by taking a more physical point of view, the smallness of $\mu$
might be seen as a requirement of linear response theory, which then naturally
leads to the correlation functions in the connection (\ref{eq:prediction}).
More details on the validity range and physical meaning can be found in Refs.\
\cite{Heitmann2023, Kraft2024}.

\begin{figure}[t]
\includegraphics[width=0.9\columnwidth]{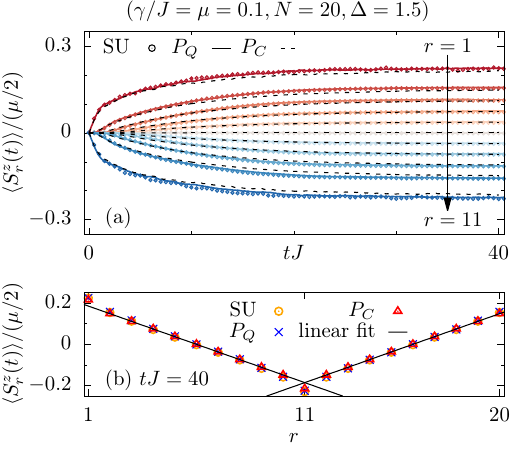}
\caption{Open-system dynamics for the model $H'$ in Eq.\
(\ref{eq:Hamiltonian_spin_perturbation}), as obtained numerically for $\Delta =
1.5$, $\Delta'=0$, $N = 20$, periodic boundary conditions, small coupling
$\gamma/J = 0.1$, and weak driving $\mu = 0.1$.
A simulation based on stochastic unraveling (SU) \cite{Heitmann2023} is
compared to two predictions, which are based on spatio-temporal correlation
functions in the closed quantum system ($P_{Q}$) \cite{Heitmann2023} and
classical system ($P_{C}$). (a) Time evolution of the local magnetization
$\langle S_{r}^{z}(t)\rangle$ for different sites $r$. (b) Site dependence of
the steady state at $tJ=40$.}
\label{fig:open_dynamics}
\end{figure}

In Fig.\ \ref{fig:open_dynamics}, we illustrate the quality of the connection,
by comparing the prediction on the r.h.s.\ of Eq.\ (\ref{eq:prediction}) to a
simulation based on stochastic unraveling for the l.h.s.\ of Eq.\
(\ref{eq:prediction}). We do so for the model $H'$ in Eq.\
(\ref{eq:Hamiltonian_spin_perturbation}) with $\Delta = 1.5$, $\Delta'=0$, $N =
20$, periodic boundary conditions, small coupling $\gamma/J = 0.1$, and weak
driving $\mu = 0.1$. For this set of parameters, the agreement between
both sides is remarkable. (Note that stochastic
unraveling as such neither requires small $\gamma$ nor weak $\mu$.) While Fig.\
\ref{fig:open_dynamics} shows existing
data from the literature \cite{Heitmann2023}, it already depicts a prediction
by the use of classical mechanics, as a main result of our work. This
prediction is discussed in the following.


\section{Classical limit} \label{sec:classical}

To introduce the classical limit of our models, let us consider an
arbitrary spin quantum number $S$. Then, the spin-$S$ operators fulfill the
usual commutation relations, which read
\begin{equation} \label{eq:commuator_relations}
 [S_{r}^{i},S_{r'}^{j}] = i \hbar \, \delta_{rr'} \sum_k \epsilon_{i j k}
\, S_{r}^{k}
\end{equation}
with the Kronecker symbol $\delta_{rr'}$ and the antisymmetric
Levi-Civita tensor $\epsilon_{i j k}$. Here, we write $\hbar$ explicitly, while
it is otherwise set to unity. The classical counterpart of our models now
results by taking the limit \cite{semiclassical}
\begin{equation} \label{eq:classical_limit}
S \to \infty \, , \quad \hbar \to 0 \, , \quad \hbar
\sqrt{S (S+1)} = 1 \, .
\end{equation}
In this limit, the commutation relations in Eq.\ (\ref{eq:commuator_relations})
turn into the Poisson-bracket relations
\begin{equation} \label{eq:Poisson_relations}
\{ S_{r}^{i},S_{r'}^{j} \} = \delta_{rr'} \sum_k \epsilon_{i j k} S_{r}^{k}
\end{equation}
for real and three-dimensional spin vectors $\mathbf{S}_{r}$ of
unit length, $|\mathbf{S}_{r}| = 1$. The Hamilton operators $H$ and $H'$
in Eqs.\ (\ref{eq:Hamiltonian_spin}) and
(\ref{eq:Hamiltonian_spin_perturbation}) become Hamilton functions, but apart
from that look the same. The total magnetization $S^z$ is still strictly
conserved, $\{ H, S^z \} = \{ H', S^z \} = 0$.

\begin{figure}[t]
\includegraphics[width=0.9\columnwidth]{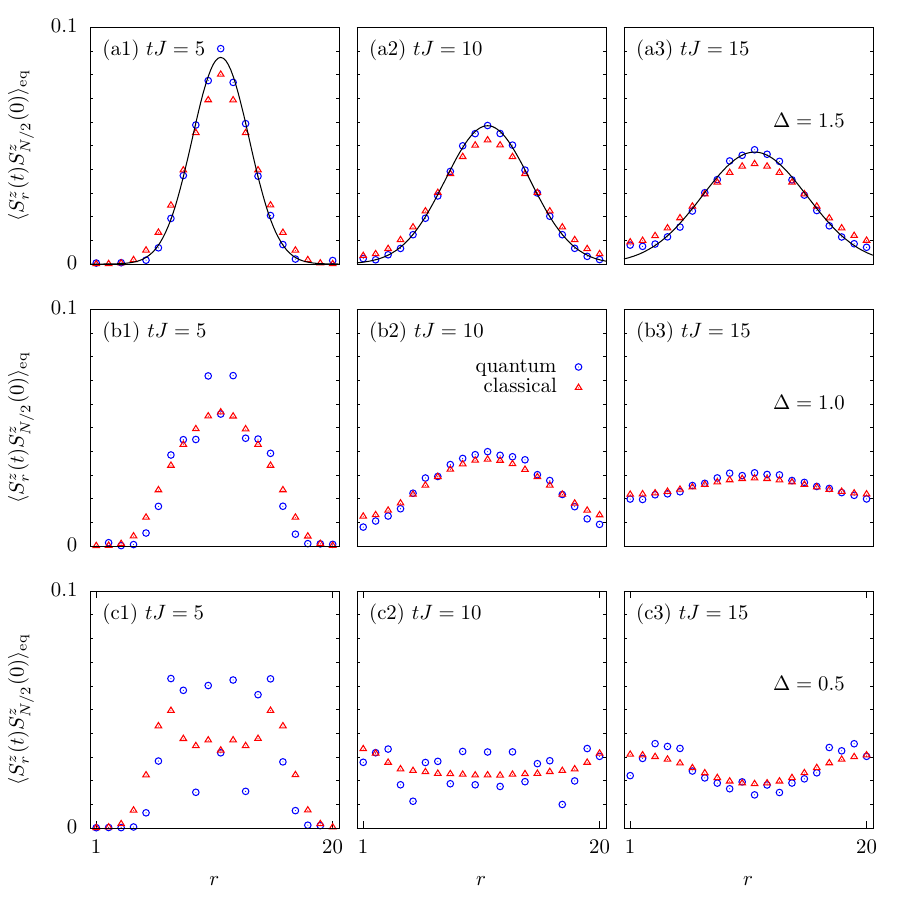}
\caption{Comparison of the site dependence of the quantum and classical
correlation function $\langle S_{r}^{z}(t) S_{r'}^{z}(0)\rangle_{\text{eq}}$ for
times $t J = 5$ (first column), $t J = 10$ (second column), and $t J = 15$
(third column) for the model $H'$ in Eq.\
(\ref{eq:Hamiltonian_spin_perturbation}) with $\Delta'=0$, $N = 20$, and
periodic boundary conditions. The anisotropy  is $\Delta = 1.5$ (first row),
$\Delta = 1.0$ (second row), and $\Delta = 0.5$ (third row). For $\Delta =
1.5$, Gaussian fits are indicated.}
\label{fig:cor_func_D}
\end{figure}

The Poisson-bracket relations in Eq.\ (\ref{eq:Poisson_relations}) also lead to
Hamilton's equations of motion,
\begin{equation} \label{eq:Hamilton_equation_of_motion}
\frac{\text{d}}{\text{d}t} \mathbf{S}_{r} = \frac{\partial H}{\partial
\mathbf{S}_{r}} \times \mathbf{S}_{r} \, ,
\end{equation}
\begin{equation}
\frac{\partial H}{\partial \mathbf{S}_{r}} =
\left(
\begin{array}{c}
S_{r-1}^x + S_{r+1}^x\\
S_{r-1}^y + S_{r+1}^y\\
\Delta (S_{r-1}^z + S_{r+1}^z) + \Delta' (S_{r-2}^z + S_{r+2}^z) \\
\end{array}
\right) \, .
\end{equation}
Physically, these equations describe the precession of a spin around a magnetic
field, which is generated by the interaction with the neighboring spins.
Mathematically, they are a coupled set of nonlinear differential equations,
which is nonintegrable by means of the Liouville-Arnold theorem, even for
$\Delta' = 0$ \cite{Arnold1978, steinigeweg2009d}, 
which is different to the quantum case $S = 1/2$. Therefore, for nontrivial
initial conditions, they have to be solved numerically, as we also do here.
Throughout our work, we employ a fourth-order Runge-Kutta scheme with a time
step $\delta t J = 0.01$, which is small enough to ensure that the total
magnetization is well conserved during the time evolution.

Now, we come to the central objects in our work, i.e., the spatio-temporal
correlation functions in the realm of classical mechanics. Focusing again on
high temperatures $\beta = 1/T \to 0$, they read
\begin{equation} \label{eq:classical_correlation_functions}
\langle S_{r}^{z}(t) S_{r'}^{z}(0) \rangle_{\text{eq}} =
\frac{1}{R_{\text{max}}} \sum_{R=1}^{R_{\text{max}}} S_{r}^{z}(t)S^{z}_{r'}(0)
\, ,
\end{equation}
where $R_\text{max}$ is the number of realizations for different initial
conditions, which are randomly drawn from a uniform
distribution of points on a unit sphere. Formally, $R_\text{max}
\to \infty$. In praxis, we average over as many realizations as $R_\text{max}
= {\cal O}(10^8 - 10^9)$, to ensure that the remaining stochastic fluctuations
are low. The need for such an extensive averaging in the numerical simulations
is kind of compensated by the fact that the phase space grows only linearly
with $N$, in contrast to the exponential growth of the Hilbert space. Due
to this fact, we can particularly treat classical systems of quite large size
$N$, which lie outside the possibilities of a quantum mechanical treatment.
Still, we also consider small $N$ to enable a comparison to quantum
mechanics for the same size.

\begin{figure}[t]
\includegraphics[width=0.9\columnwidth]{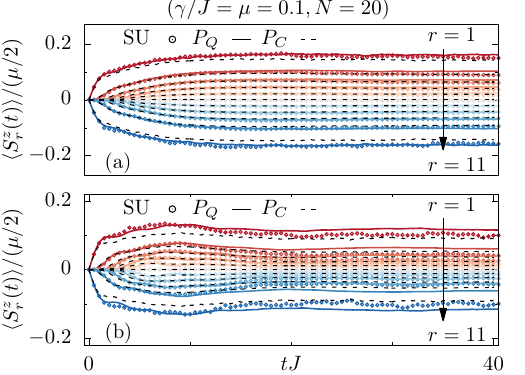}
\caption{Analogous data as the one in Fig.\ \ref{fig:open_dynamics} (a), but now
for (a) $\Delta = 1.0$ and (b) $\Delta = 0.5$. Quantum data are taken from
Ref.\ \cite{Heitmann2023}.}
\label{fig:classical_D05_D10}
\end{figure}

Eventually, coming back to the connection between the dynamics in open and
closed systems, Eq.\ (\ref{eq:prediction}), the main idea is to replace the
quantum correlation functions by the corresponding classical ones. When we do
this strong simplification, the key question of our work is whether or not the
time evolution in the open system can be still described accurately.
To answer this question in a meaningful way, we first need to
rescale the classical time by a factor
\begin{equation}
\tilde{S} = \sqrt{S(S+1)} \, ,
\end{equation}
which is $\tilde{S} \approx 0.87$ and close to one. Then, we additionally need
to rescale the initial value of the classical correlation ($\langle S_{r}^{z}(0)
S_{r'}^{z}(0) \rangle_{\text{eq}} = \delta_{rr'}/3$), since the one of the
quantum correlation ($\langle S_{r}^{z}(0) S_{r'}^{z}(0) \rangle_{\text{eq}} =
\delta_{rr'}/4$) is different. Apart from these rescalings, no further
modifications are done and the dynamics as such is fully generated by
Hamilton's equation of motion.


\section{Results} \label{sec:results}

\subsection{Nearest-neighbor interactions}

\begin{figure}[t]
\includegraphics[width=0.9\columnwidth]{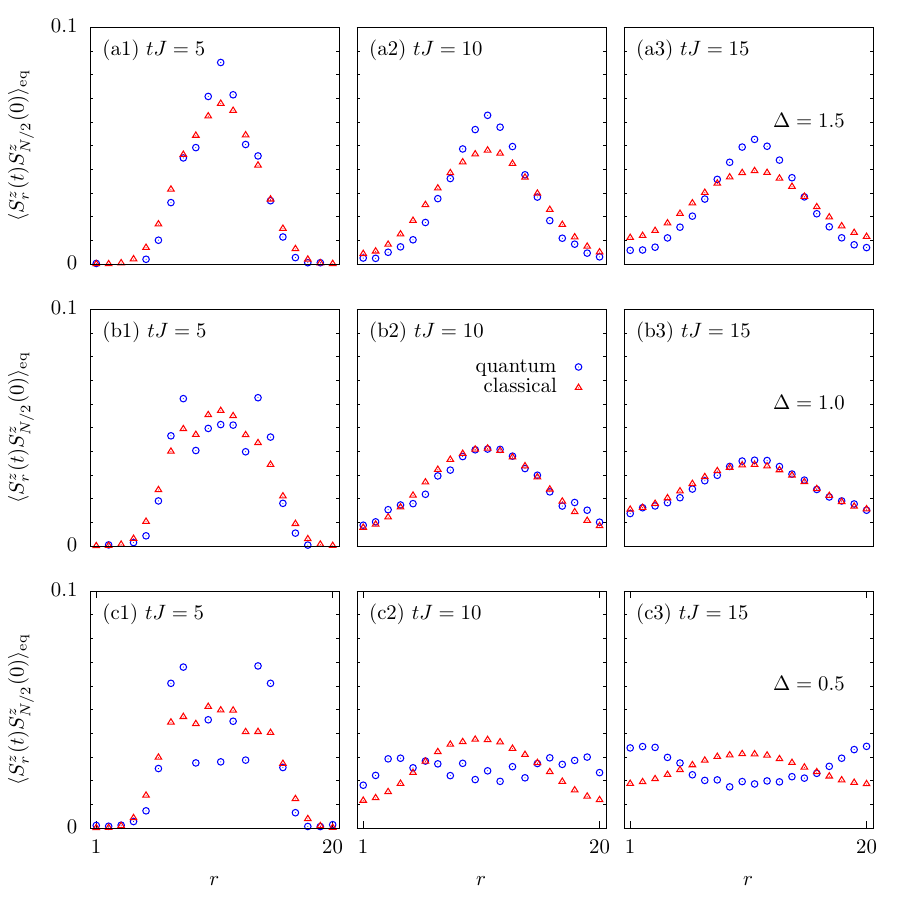}
\caption{The same data as the one in Fig.\ \ref{fig:cor_func_D}, but now with
the next-to-nearest-neighbor interaction $\Delta' = 0.5$.}
\label{fig:cor_func_D_DTwo}
\end{figure}

Next, we turn to our numerical simulations, with the central aim to scrutinize
to what degree the open-system quantum dynamics can be described by an
approach on the basis of classical mechanics. Because we rely on the connection
in Eq.\ (\ref{eq:prediction}), where correlation functions in the closed
system are the main ingredient, we start with comparing quantum and classical
correlation functions in the closed scenario. We do so for the spin-$1/2$ XXZ
chain in Eq.\ (\ref{eq:Hamiltonian_spin_perturbation}) with $\Delta' = 0$ and
$N = 20$. However, we also address the case of $\Delta' \neq 0$ and $N > 20$
later. The remaining parameters of the model are chosen to cover
different quantum behaviors \cite{Bertini2021}: (a) diffusive behavior for
$\Delta = 1.5$, (b) superdiffusive behavior for $\Delta = 1.0$, and (c)
ballistic behavior for $\Delta = 0.5$. While our calculation of quantum
correlations uses the concept of dynamical quantum typicality in its standard
formulation \cite{Heitmann2020, Jin2021}, our calculation of classical
correlations is done in the way as outlined above. Similar comparisons can be
found in the literature \cite{Heitmann2022a}.

In Fig.\ \ref{fig:cor_func_D}, we summarize the comparison, by showing the site
dependence of quantum and classical correlations for different times.
As visible in Fig.\ \ref{fig:cor_func_D} (first and second row), there is a
quite convincing agreement for $\Delta = 1.5$ and $1.0$. While minor
deviations occur at short times, the overall site dependence is similar for
longer times, which indicates that diffusion as such is not a prerequisite.
Less convincing is the comparison for $\Delta = 0.5$. Still,
quantum and classical correlations agree roughly and one might be tempted
to conclude that the transport behavior is the same. However, the quantum case
is well-known to be ballistic \cite{Bertini2021}, while the classical case is
likely diffusive, due to nonintegrability. Thus, the rough agreement reflects
that the classical dynamics has still not reached the mean free path.

\begin{figure}[t]
\includegraphics[width=0.9\columnwidth]{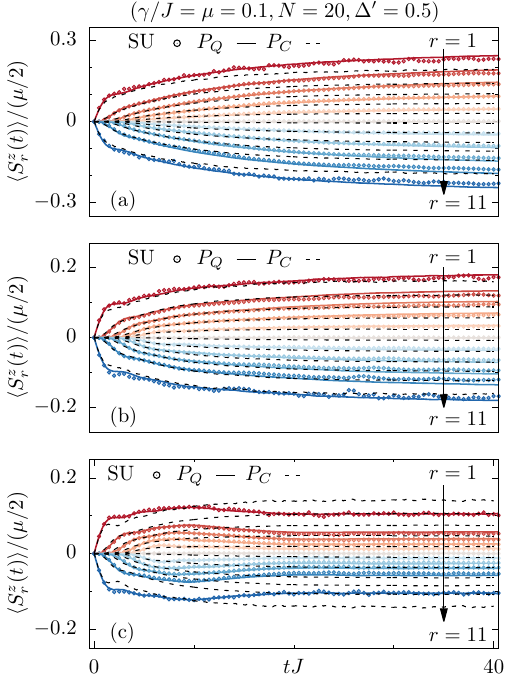}
\caption{The same data as the one in Figs.\
\ref{fig:open_dynamics} (a) and \ref{fig:classical_D05_D10}, but now with
the next-to-nearest-neighbor interaction $\Delta' = 0.5$. (a) $\Delta = 1.5$,
(b) $\Delta = 1.0$, and (c) $\Delta = 0.5$. Quantum data are taken from Ref.\
\cite{Kraft2024}.}
\label{fig:classical_D_DTwo}
\end{figure}

Using the correlations in Fig.\ \ref{fig:cor_func_D}, we now calculate the
corresponding predictions for the time evolution of the open system, according
to Eq.\ (\ref{eq:prediction}). In this calculation, we average over
$T_\text{max} = {\cal O}(10^4)$ different time sequences, which is enough to
obtain sufficiently smooth curves. As already advertised before, Fig.\
\ref{fig:open_dynamics} illustrates for $\Delta = 1.5$ a convincing agreement
with existing simulations \cite{Heitmann2023} on the basis of stochastic
unraveling. In Fig.\ \ref{fig:classical_D05_D10}, we also depict a comparison
for $\Delta = 1.0$ and $0.5$. While the agreement for $\Delta = 1.0$ is also
good, some disagreement is visible for $\Delta = 0.5$. Of course, one might
expect this disagreement in view of the differences between quantum and
classical correlations in Fig.\ \ref{fig:cor_func_D} (third row). However, it
should be noted that also the quantum prediction differs from the
stochastic-unraveling data. This deviation can be traced back
to the fact that the ballistic quantum motion for $\Delta = 0.5$ partially
violates the equilibration assumption, which underlies the derivation of Eq.\
(\ref{eq:prediction}). In other words, while the injected
excitation moves away from the boundary, it does not spread homogenoeously over
the system \cite{Heitmann2023}.

\begin{figure}[t]
\includegraphics[width=0.9\columnwidth]{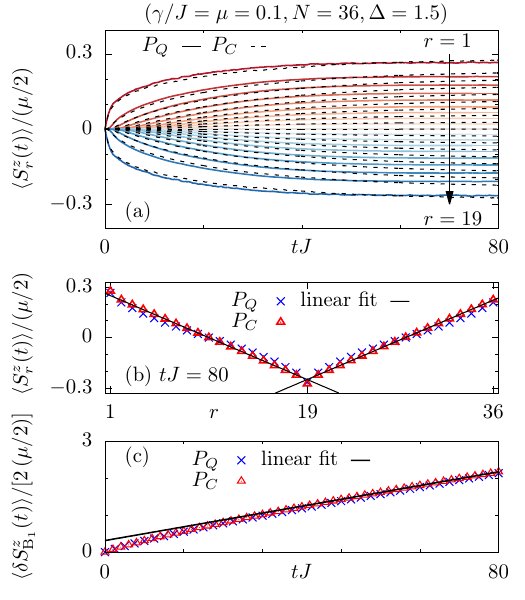}
\caption{(a) and (b) Similar data as the one in Fig.\ \ref{fig:open_dynamics},
but now for $N = 36$, where stochastic unraveling is not feasible any more. (c)
Magnetization injected by the first bath as a function of time. Quantum data
are taken from Ref.\ \cite{Heitmann2023}.}
\label{fig:open_dyn_L36}
\end{figure}

\subsection{Next-to-nearest-neighbor interactions}

Now, we allow for interactions between next-to-nearest sites and specifically
choose a value of $\Delta' = 0.5$. Naively, one might expect that the overall
agreement is improved, because the quantum and classical model become both
nonintegrable for $\Delta' \neq 0$ and as a consequence should possess
diffusion. However, this expectation turns out to be wrong.

In comparison to Fig.\ \ref{fig:cor_func_D} for $\Delta' = 0$, the quantum and
classical correlations in Fig.\ \ref{fig:cor_func_D_DTwo} for $\Delta' \neq 0$
do not agree as well as before. While the agreement for $\Delta = 1.0$ is again
convincing, deviations start to set in for $\Delta = 1.5$. These deviations
might be a signal that the quantum-classical correspondence eventually breaks
down completely in the limit of very strong interactions. Yet, the
correspondence is still satisfactory.

For $\Delta = \Delta' = 0.5$ in Fig.\ \ref{fig:cor_func_D_DTwo} (third row), the
agreement turns out to be worst. While the quantum and classical case are both
expected to exhibit diffusion, the quantum dynamics has not reached the mean
free path, at least for the time scales considered. In contrast, the classical
dynamics has reached the mean free path and already takes place
in the hydrodynamic regime. Therefore, while transport might be qualitatively
the same, it is quantitatively clearly different.

Using the correlations in Fig.\ \ref{fig:cor_func_D_DTwo}, we again calculate
the corresponding predictions for the time evolution of the open system.
As can be seen, the quality of agreement in the closed system carries over to a
similar quality in the open system. It is worth pointing out that the quantum
prediction, in contrast to the classical prediction, is in excellent
agreement with existing stochastic-unraveling data \cite{Kraft2024}, for all
values of $\Delta$.

\begin{figure}[t]
\includegraphics[width=0.9\columnwidth]{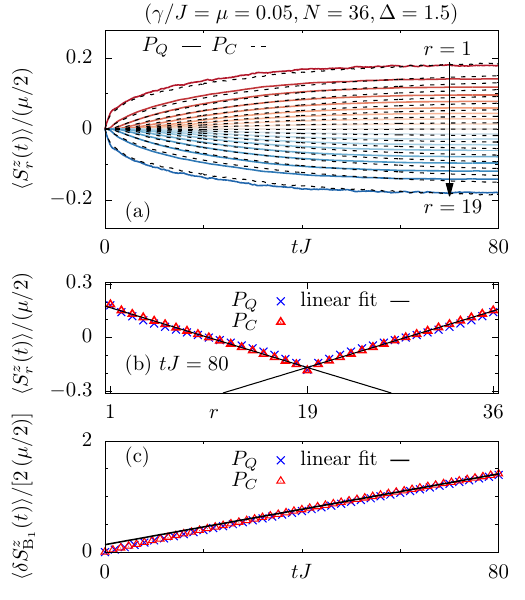}
\caption{Similar data as the one in Fig.\ \ref{fig:open_dyn_L36}, but now for
different Lindblad parameters, i.e., for a smaller system-bath coupling
$\gamma/J = 0.05$ and a weaker driving strength $\mu = 0.05$.
}
\label{fig:open_dyn_L36_G005}
\end{figure}

\subsection{Larger system sizes}

Next, we move forward to substantially larger system sizes. Here,
stochastic unraveling is not feasible anymore and we cannot compare to
corresponding data. Thus, we have to rely on a comparison between the
quantum and classical predictions for the time evolution of the open system, as
given in Eq.\ (\ref{eq:prediction}). We focus on the parameters $\Delta =
1.5$ and $\Delta' = 0$, which in our previous comparisons has turned out to be
the best case. In Fig.\ \ref{fig:open_dyn_L36}, we depict results for $N = 36$,
where the quantum prediction can still be carried out, due to existing data
for correlations \cite{Heitmann2023} from a calculation on a super computer.
Apparently, the agreement between the two predictions for $N = 36$ is as
convincing as for $N = 20$ considered before, which indicates that system size
as such is not important for the accuracy of a classical treatment.

In Fig.\ \ref{fig:open_dyn_L36} (c), we additionally depict a quantity, which
we have not discussed so far. This quantity is the injected
magnetization by the
first bath,
\begin{equation} \label{eq:IM}
\langle\delta S^{z}_{B_{1}}(t) \rangle \approx \frac{1}{T_{\text{max}}}
\sum_{T=1}^{T_{\text{max}}} \delta d_{B_1,T}(t)
\end{equation}
with
\begin{equation}
\delta d_{B_1,T}(t) = 2 \mu \sum_{j} A_{j} \, \Theta(t-\tau_{j}) \,
\langle[S_{B_{1}}^{z}(0)]^{2} \rangle \, ,
\end{equation}
as discussed in the appendix in more detail. The injected magnetization is of
interest, since from its time derivative, $ \text{d}/\text{d}t \, \langle\delta
S^{z}_{B_{1}}(t) \rangle$, the current in the steady state and then a
quantitative value for the diffusion constant can be calculated. As can be seen
in Fig.\ \ref{fig:open_dyn_L36} (c), the predictions for $\langle\delta
S^{z}_{B_{1}}(t) \rangle$ are rather close to each
other. Nevertheless, they still differ. Hence, we obtain two different
diffusion coefficients,
\begin{equation}
D_Q / J \approx 0.99 \, , \quad D_C / J \approx 0.84 \,
. \label{eq:D_36}
\end{equation}

\begin{figure}[t]
\includegraphics[width=0.9\columnwidth]{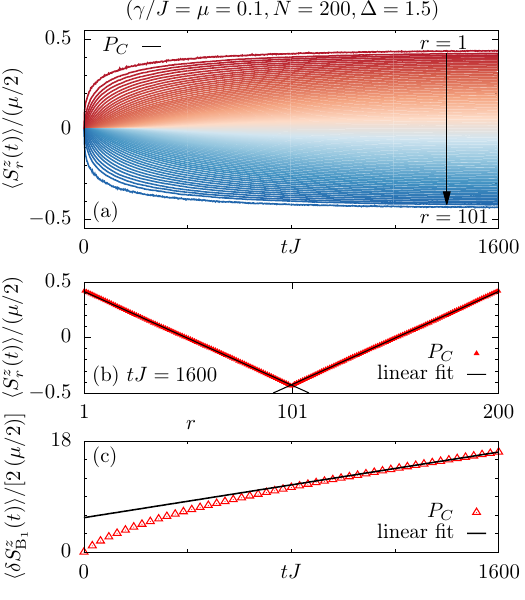}
\caption{Similar data as the one in Fig.\ \ref{fig:open_dyn_L36}, but now for a
large system size $N = 200$. For this $N$, only classical mechanics is
feasible. Note that the quality of the linear fit in (b)
is much better for $N = 200$ (compared to $N = 36$), since the relative
size of the bath-contact region is substantially smaller.}
\label{fig:open_dyn_L200}
\end{figure}

Because the prediction for the injected magnetization
could be more sensitive to short-time deviations, as
the ones visible in Fig.\ \ref{fig:cor_func_D}, we redo the analysis in Fig.\
\ref{fig:open_dyn_L36} for $\gamma/J = 0.05$ (instead of $\gamma/J = 0.1$). In
this way, the relevant time scale is increased. Simultaneously, we use another
$\mu = 0.05$ (instead of $\mu = 0.1$) in Fig.\ \ref{fig:open_dyn_L36_G005}.
From this data, however, a significant change of the
diffusion constants does not result,
\begin{equation} \label{eq:values_D}
D_Q / J \approx 1.01 \, , \quad  D_C / J \approx 0.86
\, ,
\end{equation}
which indicates that we do not need to devote special
attention to the role of the already small $\gamma$ and $\mu$, to extract
quantitative values for transport constants.

It is instructive to compare the values for the diffusion constants in Eq.\
(\ref{eq:values_D}) to other values in the literature. In the classical case, a
value of $D_C/J \sim 0.6$ has been found for the closed system
\cite{Steinigeweg2012}, which is consistent with the
value for the open system
in Eq.\ (\ref{eq:values_D}). In the quantum case, the value is
known with less precision and might be $D_Q/J \sim 0.6$ or larger in the closed
system \cite{Bertini2021}, while a value of $D_Q/J \approx 0.58$ has been found
in the open system \cite{Prosen2009, Znidaric2011}, based on
matrix-product-state simulations of the Lindblad equation, yet for a larger
$\gamma/J = 1$.

Finally, we demonstrate in Fig.\ \ref{fig:open_dyn_L200} explicitly that our
classical approach can be used for as many as $N = 200$ lattice sites, where a
quantum mechanical treatment is impossible, at least for the techniques
used by us. While the required time to reach the steady state increases with
$N$, the additional computing time poses no conceptual problem. This result is
particularly relevant, since it is obtained for a small $\gamma$, which is hard
to treat on the basis of matrix product states. For the diffusion constant, we
find
\begin{equation}
D_C / J \approx 0.79 \, ,
\end{equation}
which is close to $D_C / J \approx 0.84$ for $N = 36$,
cf.\ Eq.\ (\ref{eq:D_36}), and indicates a good convergence w.r.t.\ system
size.

\begin{figure}[t]
\includegraphics[width=0.9\columnwidth]{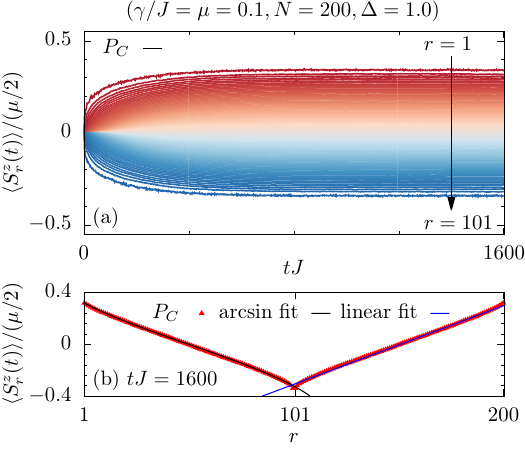}
\caption{Classical data for a large system size $N = 200$ and the isotropic
point $\Delta = 1.0$. The other model parameters are the same as the ones in
Fig.\ \ref{fig:open_dyn_L200}. For the steady-state profile in (b), a fit with
the function $f(r) = a \, \arcsin[b \, (r-L/4)/(L/4)]$ is shown
\cite{Znidaric2011}, in addition to a linear fit.}
\label{fig:open_dyn_L200_Delta1}
\end{figure}

\begin{figure}[b]
\includegraphics[width=0.9\columnwidth]{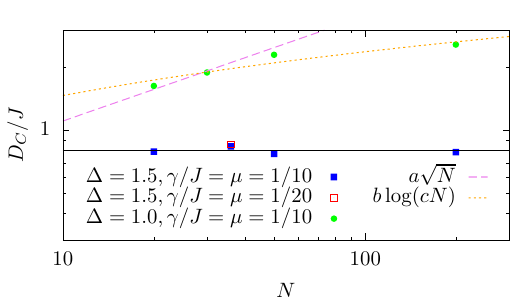}
\caption{Classical data for the system-size dependence
of the diffusion constant $D$. While $D$ does not change significantly with
$N$ for $\Delta = 1.5$, it increases with $N$ for $\Delta = 1.0$, which
indicates superdiffusion \cite{Znidaric2011}. A power law and a logarithm are
depicted as a guide to the eye.}
\label{fig:current_scaling}
\end{figure}

Since Fig.\ \ref{fig:open_dyn_L200} focuses on $\Delta = 1.5$, we
depict in Fig.\ \ref{fig:open_dyn_L200_Delta1} additional $N = 200$ data for
$\Delta = 1.0$. In comparison to $\Delta = 1.5$, the steady-state profile for
$\Delta = 1.0$ is not as well described by a linear function. Thus, we perform
in Fig.\ \ref{fig:open_dyn_L200_Delta1} (b) a fit with another function,
\begin{equation}
f(r) = a \, \arcsin \Big [ \frac{b \, (r-L/4)}{L/4} \Big ] \, ,
\end{equation}
which has been used in the quantum case $S = 1/2$ before \cite{Znidaric2011}.
And indeed, also the classical case is well described by such a function, which
indicates the usefulness of our classical approach for situations beyond
the one of normal diffusion.

To further substantiate this claim and to conclude on
the specific type of anamolous transport, we also depict in Fig.\
\ref{fig:current_scaling} classical data for the system-size dependence
of the diffusion constant $D$. At the isotropic point $\Delta = 1.0$, $D$
increases with $N$ and indicates superdiffusion \cite{Znidaric2011}. In
contrast, at $\Delta = 1.5$, $D$ does not change significantly with
$N$, as expected for diffusion. While we do not know the specific scaling
function for $\Delta = 1.0$, we depict in Fig.\
\ref{fig:current_scaling} a power law and a logarithm as a guide
to the eye. We should also stress that we cannot exclude a saturation
in the large-system limit. For an early discussion of this topic, see Refs.\
\cite{Mueller1988, Gerling1989, AlcantaraBonfim1992, Boehm1993}.


\section{Conclusion} \label{sec:conclusion}

In summary, we have studied the Lindblad equation, as a central approach to
the physics of open quantum systems. In this context, we have particularly
focused on boundary-driven transport, where a steady state emerges in the
long-time limit, which features a constant current and a characteristic density
profile. The starting point of our study has been a recently suggested
connection \cite{Heitmann2023} between the dynamics in open and closed
systems, which exits in certain cases. We have built on this connection for
magnetization transport in the spin-1/2 XXZ chain with and without
next-to-nearest-neighbor interactions, as integrability breaking perturbations.
Specifically, we have studied the question whether the time
evolution of the open quantum system can be described by the use of classical
correlation functions, as generated by the Hamiltonian equations of motion for
classical real vectors. By comparing to exact numerical simulations of the
Lindblad equation, we have found a good accuracy of such a
description for a range of model parameters, but we have also
pointed out counterexamples and its limitations for the quantitative extraction
of transport coefficients. While this agreement is an interesting physical
observation, it has also suggested that classical mechanics can be used to
solve the Lindblad equation for comparatively large systems, which cannot be
reached in a quantum mechanical treatment. Despite the approximate nature of
this approach, one particular strength is the possibility to deal with small
system-bath couplings, which are much harder to access on the basis of other
numerical techniques.

An interesting question is the precise regime of
validity of the quantum-classical correspondence of open systems within
infinite-temperature spin-chain models. To answer it, one first has to know in
which cases quantum and classical dynamics agree in the closed setup. Previous
works like \cite{Schubert2021} suggest that such an agreement typically exists
if the quantum model is nonintegrable, as it usually is for quasi-1D or 2D
lattices. An agreement can also exist if the quantum model is integrable, as
shown in Fig.\ \ref{fig:open_dynamics}. Yet, it necessarily requires that
quasilocal conserved charges are of no relevance to the transport behavior
\cite{Bertini2021}, which is true for $\Delta > 1$ in Fig.\
\ref{fig:open_dynamics}, but not for $\Delta < 1$. An related point of view is
that for $\Delta < 1$ the system is more XY-like and quantum correlations become
important.

Another interesting question is the existence of a
quantum-classical correspondence of open systems outside infinite temperature
and spin models. Since our present work relies on the relation in and around
Eq.\ (\ref{eq:prediction}), one first has to answer if this relation can be
generalized to other models and finite temperatures, which is a promising
direction of future research. Apart from the generalization of Eq.\
(\ref{eq:prediction}), it is clear that quantum and classical dynamics cannot
agree for all models and finite temperatures, neither in the closed nor in the
open setup. Already for spin systems, the nature of low-energy excitations can
be different in the quantum and classical case. Moreover, some systems have no
classical counterpart.


\subsection*{Acknowledgments}

We thank Jochen Gemmer for fruitful discussions. This work has been funded by
the Deutsche Forschungsgemeinschaft (DFG), under Grant No.\ 531128043, as well
as under Grant No.\ 397107022, No.\ 397067869, and No.\ 397082825 within the
DFG Research Unit FOR 2692, under Grant No.\ 355031190.

\appendix*


\section{Current in the steady state}

In addition to the dynamics of the local magnetization, one can predict the
current in the  steady state. To this end, one needs the injected magnetization
$\langle \delta S^{z}_{B_{1}} \rangle$ at the bath site $B_{1}$, where the
symbol $\delta$ is just a notation for the term ``injected''. This injected
magnetization can be predicted as \cite{Heitmann2023}
\begin{equation}
\langle\delta S^{z}_{B_{1}}(t) \rangle \approx \frac{1}{T_{\text{max}}}
\sum_{T=1}^{T_{\text{max}}} \delta d_{B_1,T}(t)
\end{equation}
with
\begin{equation}
\delta d_{B_1,T}(t) = 2 \mu \sum_{j} A_{j} \, \Theta(t-\tau_{j}) \,
\langle[S_{B_{1}}^{z}(0)]^{2} \rangle \, ,
\end{equation}
which is slightly simpler than Eq.\ (\ref{eq:prediction}) in the main text and
can be related to the current of interest. Since in the steady state all local
currents are the same,
\begin{equation}
\langle j_r \rangle = \langle j_{r'}\rangle, \;\;\; B_{1} \leq r,r' \leq B_{2}
\, ,
\end{equation}
one only needs to know $\langle j_{B_{1}}\rangle$, which can be
calculated as
\begin{eqnarray}\label{eq:current}
 \langle j_{B_{1}} \rangle = \frac{\text{d}}{\text{d}t} \frac{\langle \delta
S_{B_{1}}^{z} (t) \rangle }{2} \; .
\end{eqnarray}
Note that the factor $2$ occurring in the denominator is due to periodic
boundary conditions, as magnetization can flow to the right and left of the bath
contact.

The diffusion constant follows from $\langle j_{B_{1}} \rangle$ via
\begin{equation} \label{eq:diffusion_constant}
 D = -\frac{\langle j_{B_{1}}\rangle}{\langle S_{r+1}^{z} \rangle - \langle
S_{r}^{z}\rangle}
\end{equation}
for some site $r$ in the bulk of the system.



\begin{thebibliography}{81}
\bibitem [{SU1()}]{SU1}%
  \BibitemOpen
  \href@noop {} {}\bibinfo {note} {The Lindblad description has
  a weak U(1) symmetry, which might be used to simplify exact
  diagonalization.}\BibitemShut {Stop}%
\bibitem [{sem()}]{semiclassical}%
  \BibitemOpen
  \href@noop {} {}\bibinfo {note} {In this work, we use only
  the limit in Eq.\ (\ref{eq:classical_limit}) and no additional semiclassical
  approximation.}\BibitemShut {Stop}%
\end{thebibliography}

%

\clearpage
\newpage

\end{document}